\documentstyle[preprint,prl,tighten,aps,epsfig]{revtex} 

\begin{document}

\draft

\preprint{\begin{tabular}{l}
\hbox to\hsize{\mbox{ }\hfill KIAS-P99077}\\
\hbox to\hsize{\mbox{ }\hfill \today}\\
\hbox to\hsize{\mbox{ }\hfill { } }\\
\hbox to\hsize{\mbox{ }\hfill { } }\\
          \end{tabular} }

\title{Neutralino--Nucleus Elastic Scattering\\
       in the MSSM with Explicit CP Violation}

\author{S.Y.~Choi}
\address{Korea Institute for Advanced Study (KIAS), 207--43, 
         Cheongryangri--dong, \\
         Dongdaemun--gu, Seoul 130--012, Korea}

\maketitle

\vskip 2cm
\begin{abstract}
We re--investigate the neutralino--nucleus elastic scattering as a 
promising dark matter detection mechanism including contributions from 
the scalar--pseudoscalar mixing of neutral Higgs states and the induced 
phase between two Higgs doublets due to the CP--violating phases
of the scalar top and bottom sectors in the minimal supersymmetric 
standard model. The spin--dependent part of the cross section
turns out to be hardly affected by the CP--violating induced phase due to a
mutually destructive unavoidable suppression mechanism of various relevant 
supersymmetric parameters. On the other hand, although the phase $\Phi_\mu$ 
of the higgsino mass parameter is set to be zero, the spin--independent part, 
which can dominate over the spin--dependent part for heavy nuclei, can be 
strongly dependent on the CP--violating scalar--pseudoscalar
mixing and the induced phase, in particular, for a large $|\mu|$, 
a small charged Higgs boson mass, and a large trilinear term $|A|$
compared to the SUSY breaking scale. For a small $\tan\beta$ and a 
large $|\mu|$, the spin--independent cross section 
is enhanced by an order of magnitude as the phase $\Phi_A$ of the
trilinear term increases up to $\pi$, while for a large $\tan\beta$ 
and a large $|\mu|$ the spin--independent cross section is 
significantly suppressed for non--zero values of the phase $\Phi_A$.
\end{abstract}
\vskip 0.4cm



\section{Introduction}

Supersymmetry (SUSY) is one of the most promising theoretical
frameworks for a successful unification of gravity with 
all other fundamental forces and is the most appealing perturbative 
solution to the gauge hierarchy problem of the standard model (SM). 
However, SUSY is not an exact symmetry 
of nature. The minimal supersymmetric SM (MSSM) \cite{MSSM},
the minimal SUSY realization, must break SUSY softly in order to 
accomplish agreement with experimental observations and its  
breaking scale should not be much larger than a few TeV in order
to retain  naturalness.
In general, the SUSY breakdown introduces a large number 
of unknown parameters, many of which can be complex \cite{MS}. 
CP--violating phases associated with sfermions of the first and, to
a lesser extent, second generations are severely constrained by bounds
on the electric dipole moments of the electron, neutron and muon.
The present experimental upper bounds on the neutron EDM $d_n$ 
and electron EDM $d_e$ are  very tight \cite{eEDM}: 
$|d_n|< 1.12\times 10^{-25}\, e\, {\rm cm}$ and $|d_e|< 0.5\times
10^{-26}\, e\,{\rm cm}$ at the 2--$\sigma$ level.
As a result of the CP crises, some fine--tuning mechanisms
are necessary in generic supersymmetric theories to avoid these problems.
There have been several phenomenologically attractive solutions 
\cite{KO,Kaplan,IN} to evade these constraints  without suppressing 
the CP--violating phases. One option is to make the first two generations 
of scalar fermions rather heavy so that one--loop EDM constraints are 
automatically evaded. As a matter of fact one can consider so--called 
effective SUSY models \cite{Kaplan} which seem to combine all healthy 
features of both the MSSM and technicolor theories. The main virtue of
the effective SUSY model is that any non--SM source of CP violation and
FCNC involving the first two generations is suppressed by allowing their
respective soft--SUSY--breaking masses to be as high as 20 TeV, whereas
third generation scalar quarks and leptons may naturally be light well
below the TeV scale. Another possibility is to arrange for partial 
cancellations among various contributions to the electron and neutron 
EDM's \cite{IN}.

Following the suggestions that the CP--violating phases do not have to 
be always suppressed, many important works on the effects due to the 
CP--violating phases in the MSSM have been already reported; 
the effects are very significant in extracting the parameters in the 
SUSY Lagrangian from experimental data \cite{SYCHOI}, 
estimating dark matter densities and scattering cross sections and 
Higgs boson mass limits \cite{BK,FO,PW}, CP violation in the $B$ and $K$ 
systems \cite{Ko}, and so on. In particular, it has been found \cite{PW} 
that the Higgs--sector CP violation induced via loop corrections of
soft CP--violating Yukawa interactions may drastically modify the 
couplings of the light neutral Higgs boson to the gauge bosons.
As a result, the production cross sections as well as the decay 
branching ratios \cite{CL} of neutral Higgs bosons changes so
significantly. One of the crucial modifications is that the current 
experimental lower bound on the lightest Higgs boson mass may be 
dramatically relaxed up to a 60--GeV level in the presence of large 
CP violation in the Higgs sector of the MSSM. Moreover, the explicit 
CP violation in the MSSM Higgs sector radiatively induces a finite 
unremovable misalignment \cite{DEMIR} between two Higgs doublets.
This additional phase can be as large as the original CP phases in certain
portions of the MSSM parameter space and affect the chargino and neutralino
systems. Therefore, it would be very important to re--visit all the relevant
phenomena by including the CP--violating induced phase while avoiding the severe 
constraints from the neutron and electron EDMs. In particular, one of the most 
phenomenologically interesting subjects is to investigate the possibility
of detecting lightest neutralinos when a large portion of the
dark matter in the Universe is composed of the lightest and (almost) stable
lightest supersymmetric particle (LSP).

The observations of the dynamics of galaxies and 
clusters of galaxies \cite{GDYN}, and the constraints on the 
baryon density from big bang nucleosynthesis \cite{BBN} requires 
the existence of a considerable amount of non-baryonic dark matter. 
Therefore, it is almost universally accepted that 
most of the mass in the Universe and most of the mass in the Galactic 
halo is dark and the dark matter consists of some new, as yet undiscovered, 
weakly--interacting massive particle (WIMP). 
Of the many WIMP candidates, one of the best motivated and 
the most theoretically developed is the lightest neutralino, the 
lightest supersymmetric particle (LSP) in most SUSY theories. 
In this light, there have been an intensive investigation for 
its detection and identification \cite{JKG}. 
In the present work, we re--investigate the LSP--nucleus 
elastic scattering process in the MSSM framework with R-parity and 
with CP--violating complex parameters. There have been already several 
works \cite{FO} on the effects of the phase $\Phi_\mu$ of the 
higgsino mass parameter $\mu$ on the neutralino--nucleus elastic 
scattering as well as the neutralino relic density \cite{FOS}. 
So, referring to those works for the effects from the 
phase $\Phi_\mu$, we will mainly concentrate on the impact of the 
scalar--pseudoscalar mixing and the induced phase between two Higgs 
doublets in the MSSM Higgs sector stemming from the radiative corrections 
due to the CP--violating scalar top and bottom 
sectors in the MSSM on the neutralino--nucleus elastic scattering.
For a more concrete, quantitative investigation, our analysis through
the paper is based on a specific scenario with the following assumptions:
\begin{itemize}
\item The first and second generation sfermions are very heavy so that 
      they are decoupled from the theory. In this case, there are no
      constraints on the CP--violating phases from the neutron and electron
      EDMs. On the other hand, the annihilation of the neutralinos into
      tau pairs through the exchange for relatively light scalar tau leptons
      guarantees that the cosmological constraints on the dark matter densities 
      be satisfied.
\item The explicit CP violation in the Higgs sector through the CP--violating
      radiative corrections from the scalar top and bottom sectors is 
      included.
\item Simultaneously, the effects of the induced CP--violating phase between
      two Higgs doublets on the chargino and neutralino systems 
      are explicitly included.
\item It is necessary to avoid the possible constraints from the so--called 
      Barr-Zee--type diagrams \cite{Pilaftsis} to the electron and neutron 
      EDMs as well as from the null results of the Higgs boson searches at 
      LEP \cite{HIGGS}. We take two values 3 and 30 for $\tan\beta$, the ratio 
      of the vacuum expectation values of two neutral Higgs fields in the 
      present analysis.
\end{itemize}

Certainly, the major issue concerning the supersymmetric dark matter 
is its detection and identification.  Indeed, there are a multitude of
ongoing experiments involved in the direct and indirect detection of
dark matter, many with a specific emphasis on searching for
supersymmetric dark matter \cite{JKG}.  The event rates for  either
direct or indirect detection depend crucially on the LSP-nucleon, or 
LSP-nucleus, cross-section.  Because the neutralinos have Majorana 
mass terms, their interactions with matter are generally spin dependent, 
coming from an effective interaction term of the form 
$(\bar{\chi}\gamma^\mu\gamma_5\chi)(\bar{f}\gamma_\mu\gamma_5 f)$.  
In the regions of the MSSM 
parameter space where the LSP is a mixture of both gaugino and Higgsino 
components, there is also an important contribution to the scattering
cross-section due to a term in the interaction Lagrangian of the form 
$(\bar{\chi}\chi)(\bar{f} f)$ \cite{G} which is spin independent.  These
terms are particularly important for scattering off of large nuclei,
where coherent nucleon scattering effects can quickly come to dominate
all others. The scalar--pseudoscalar mixing and the modified couplings
of the neutral Higgs bosons to fermions and neutralinos affect 
the spin--independent part while the CP--violating induced phase 
affects both the spin--dependent and the spin--independent parts
through its modifications of the structure of the neutralino mass matrix.  
In this light it is worthwhile to make a systematic investigation of the
effects of the CP--violating phases on the neutralino--nucleus 
scattering cross section, which is the goal of the present work.

The organization of the present paper is as follows.
In Section II, a brief review on the explicit CP violation in the Higgs
sector is given following the work by Pilaftsis and Wager \cite{PW}.
Section III is devoted to a detailed analysis of various effects of the 
CP--violating induced phase between two Higgs doublets on the chargino and
neutralino sectors. Then we give the fully analytic expressions
for the spin--dependent and spin--independent neutralino--nucleus scattering
cross sections in Section IV and give a detailed numerical analysis of the
dependence of the cross sections on the CP--violating phases as well
as real SUSY parameters such as $\tan\beta$, the size of the higgsino 
mass parameter $|\mu|$ and the trilinear terms $|A_{t,b}|$.
Finally, we summarize our findings and conclude in Section V.

\section{CP Violation in the MSSM Higgs sector}
\label{sec:CP violation}

\subsection{CP--violating radiative corrections}

The MSSM introduces several new parameters in the theory that are absent
from the SM and could, in principle, possess many CP--violating phases. 
Specifically, the new CP phases may come from the following parameters: 
(i) the higgsino mass parameter $\mu$, which involves the bilinear mixing 
of the two Higgs chiral superfields in the superpotential; (ii) the soft 
SUSY--breaking gaugino masses $M_a$ ($a=1,2,3$), where the index $a$ 
stands for the gauge groups U(1)$_Y$, SU(2)$_L$ and SU(3)$_C$, 
respectively; (iii) the soft bilinear Higgs mixing masses $m^2_{12}$, 
which is sometimes denoted as $B\mu$ in the literature; (iv) the soft 
trilinear Yukawa couplings $A_f$ of the Higgs particles to scalar 
fermions; and (v) the flavor mixing elements of the sfermions mass matrices. 
If the universality condition is imposed on all gaugino masses 
at the unification scale $M_X$, the gaugino masses $M_a$ have a common 
phase, and if the diagonal boundary conditions are added to the
universality condition for the sfermion mass matrices at the GUT
scale, the flavor mixing elements of the sfermions mass matrices vanish 
and the different trilinear couplings $A_f$ are all equal, i.e. $A_f=A$. 

The conformal--invariant part of the MSSM Lagrangian has two 
global U(1) symmetries; the U(1)$_Q$ Peccei--Quinn symmetry
and the U(1)$_R$ symmetry acting on the Grassmann--valued coordinates.
As a consequence, not all CP--violating phases of the four complex 
parameters $\{\mu,m^2_{12},M_a,A\}$ turn out to be physical, i.e. 
two phases may be removed by redefining the fields accordingly \cite{DGH}. 
Employing the two global 
symmetries, one of the Higgs doublets and the gaugino
fields can be rephased such that $M_a$ and $m^2_{12}$ become real.
In this case, arg($\mu$) and arg($A$) are the only physical
CP--violating phases in the low--energy MSSM supplemented by universal 
boundary conditions at the GUT scale.
Denoting the scalar components of the Higgs doublets $H_1$ and $H_2$ by
$H_1=-i\tau_2 \Phi^*_1$ ($\tau_2$ is the usual Pauli matrix) and 
$H_2=\Phi_2$, the most general CP--violating 
Higgs potential of the MSSM can be conveniently described by the effective
Lagrangian
\begin{eqnarray}
{\cal L}_V&=&\mu^2_1(\Phi^\dagger_1\Phi_1)
        +\mu^2_2(\Phi^\dagger_2\Phi_2)
        +m^2_{12}(\Phi^\dagger_1\Phi_2)
        +m^{*2}_{12}(\Phi^\dagger_2\Phi_1)\nonumber\\
      &&+\lambda_1(\Phi^\dagger_1\Phi_1)^2
        +\lambda_2(\Phi^\dagger_2\Phi_2)^2
        +\lambda_3(\Phi^\dagger_1\Phi_1)(\Phi^\dagger_2\Phi_2)
        +\lambda_4(\Phi^\dagger_1\Phi_2)(\Phi^\dagger_2\Phi_1)\nonumber\\
      &&+\lambda_5(\Phi^\dagger_1\Phi_2)^2
        +\lambda^*_5(\Phi^\dagger_2\Phi_1)^2
        +\lambda_6(\Phi^\dagger_1\Phi_1)(\Phi^\dagger_1\Phi_2)
        +\lambda^*_6(\Phi^\dagger_1\Phi_1)(\Phi^\dagger_2\Phi_1)\nonumber\\
      &&+\lambda_7(\Phi^\dagger_2\Phi_2)(\Phi^\dagger_1\Phi_2)
        +\lambda^*_7(\Phi^\dagger_2\Phi_2)(\Phi^\dagger_2\Phi_1)\,.
\label{eq:Higgs potential}
\end{eqnarray}
In the Born approximation, the quartic couplings $\lambda_{1,2,3,4}$ are 
solely determined by the gauge couplings and $\lambda_{5,6,7}$ are zero. 
However, beyond the Born approximation, the quartic couplings 
$\lambda_{5,6,7}$ receive significant radiative corrections from 
trilinear Yukawa couplings of the Higgs fields to scalar--top and 
scalar--bottom quarks. These parameters are in general complex and so 
lead to CP violation in the Higgs sector through radiative corrections. 
The explicit form of the couplings with radiative corrections 
can be found in Refs.~\cite{PW,HH}.

It is necessary to determine the ground state of the Higgs potential to 
obtain physical Higgs states and their self--interactions. To this end
we introduce the linear decompositions of the Higgs fields
\begin{eqnarray}
\Phi_1=\left(\begin{array}{cc}
             \phi^+_1 \\
	     \frac{1}{\sqrt{2}}(v_1+\phi_1+ia_1)
	     \end{array}\right)\,, \qquad
\Phi_2={\rm e}^{i\xi}\left(\begin{array}{cc}
             \phi^+_2 \\
	     \frac{1}{\sqrt{2}}(v_2+\phi_2+ia_2)
	     \end{array}\right)\,,
\end{eqnarray}
with $v_1$ and $v_2$ the moduli of the vacuum expectation values (VEVs) 
of the Higgs doublets and $\xi$ their CP--violating induced relative phase. 
These VEVs and the relative phase can be determined by the minimization 
conditions on ${\cal L}_V$, which can be efficiently performed 
by the so--called tadpole renormalization techniques \cite{PW,PILL}. 
It is always guaranteed that one combination of the CP--odd  Higgs 
fields $a_1$ and $a_2$ ($G^0=\cos\beta a_1
-\sin\beta a_2$) defines a flat direction in the Higgs potential and 
so it is absorbed as the longitudinal component of the $Z$ boson. 
[Here, $\sin\beta=v_2/\sqrt{v^2_1+v^2_2}$ and 
$\cos\beta=v_1/\sqrt{v^2_1+v^2_2}$.] As a result, there exist one charged 
Higgs state and three neutral Higgs 
states that are mixed in the presence of CP violation in the Higgs 
sector. Denoting the remaining CP--odd state $a=\-\sin\beta a_1
+\cos\beta a_2$, the $3\times 3$ neutral Higgs--boson mass matrix 
describing the mixing between CP--even and CP--odd fields can be 
decomposed into four parts in the weak basis $(a,\phi_1,\phi_2)$ :
\begin{eqnarray}
{\cal M}^2_H=\left(\begin{array}{cc}
                   {\cal M}^2_P     &   {\cal M}^2_{PS} \\
		   {\cal M}^2_{SP}  &   {\cal M}^2_S 
		   \end{array}\right)\,,
\end{eqnarray}
where ${\cal M}^2_P$ and ${\cal M}^2_S$ describe the CP--preserving
transitions $a\rightarrow a$ and $(\phi_1,\phi_2)\rightarrow 
(\phi_1,\phi_2)$, respectively, and ${\cal M}^2_{PS}=({\cal M}^2_{SP})^T$ 
contains the CP--violating mixings $a\leftrightarrow (\phi_1,\phi_2)$. 
The analytic form of the sub--matrices can be found in Ref.~\cite{PW}.

On the other hand, the charged Higgs-boson mass $m_{H^\pm}$ is related
to the pseudoscalar mass term $m_a$ as
\begin{eqnarray}
m^2_a = m^2_{H^\pm}-\frac{1}{2}\lambda_4 v^2
           +{\cal R}(\lambda_5{\rm e}^{2i\xi}) v^2\,.
\end{eqnarray}
Taking this very last relation between $m_{H^\pm}$ and $m_a$ into account,
we can express the neutral
Higgs--boson masses as functions of $m_{H^\pm}$, $\mu$, $A$, 
a common SUSY scale $M_{\rm SUSY}$, $\tan\beta$ and the physical
phase $\xi$, which cannot be rotated away in the presence of  
the chargino and neutralino contributions.
Clearly, the CP--even and CP--odd states mix unless all of the
imaginary parts of the parameters $\lambda_{5,6,7}$ vanish. Since
the Higgs---boson mass matrix ${\cal M}^2_H$ describing the 
scalar--pseudoscalar mixing is 
symmetric, we can diagonalize it by means of an orthogonal rotation $O$;
$O^T{\cal M}^2_H O={\rm diag}(m^2_{H_3},m^2_{H_2},m^2_{H_1})$
with the ordering of masses $m_{H_1}\leq m_{H_2}\leq m_{H_3}$.
The neutral Higgs--boson mixing affects the couplings of the Higgs fields 
to fermions, gauge bosons, and Higgs fields themselves as shown in the
following. 

The CP--violating effects due to radiative corrections to the Higgs 
potential is characterized by a dimensionless parameter $\eta_{_CP}$
\begin{eqnarray}
\eta_{_{CP}}=\frac{m^4_{t,b}}{v^4}
             \left(\frac{|\mu||A|}{32\pi^2 M^2_{\rm SUSY}}
          \right)\sin\Phi_{_{CP}}\,,
\end{eqnarray}
where $\Phi_{_{CP}}={\rm arg}(A\mu)+\xi$, i.e. the sum of three CP--violating
phases. So, for $|\mu|$ and/or $|A|$ values larger than the 
SUSY--breaking 
scale $M_{\rm SUSY}$, the CP--violating effects can be significant.

\subsection{An induced CP--violating phase}

As shown in the previous section, the first derivatives of the Higgs
potention with respect to the neutral fields $\{\phi_1,\phi_2,a\}$ do not
vanish any longer; hence, one has to redefine the Higgs doublet fields
with a relative phase $\xi$. This CP--violating induced phase can be
obtained analytically by combining the two relations for the minimization.
First of all, we make use of the fact that a U(1)$_{PQ}$ rotation allows
us to take $m^2_{12}$ to be real and for a notational convenience define 
$\tilde{\lambda}_6$, $\delta$, and $\delta'$;
\begin{eqnarray}
&& \tilde{\lambda}_6=\lambda_6\, c^2_\beta+\lambda_7\, s^2_\beta\,,\nonumber\\
&& \delta = \left(\frac{m^2_{H^\pm}}{v^2}-\frac{\lambda_4}{2}
                 +\lambda_5\right)\sin 2\beta\,, \qquad
   \delta'= \left(\frac{m^2_{H^\pm}}{v^2}-\frac{\lambda_4}{2}
                 -\lambda_5\right)\sin 2\beta\,. 
\end{eqnarray}
Then, the CP--violating induced phase $\xi$ is determined by the relations;
\begin{eqnarray}
&& \sin\xi = -\frac{1}{|\delta|^2}\left\{ 
              {\cal R}(\delta) {\cal I}(\tilde{\lambda}_6) 
             -{\cal I}(\delta)\sqrt{|\delta|^2-{\cal I}^2(\tilde{\lambda}_6)}
                                  \right\}\,, \nonumber\\
&& \cos\xi = +\frac{1}{|\delta|^2}\left\{ 
              {\cal I}(\delta) {\cal I}(\tilde{\lambda}_6) 
             +{\cal R}(\delta)\sqrt{|\delta|^2-{\cal I}^2(\tilde{\lambda}_6)}
                                  \right\}\,, 
\end{eqnarray}
and the soft--breaking positive bilinear mass squared $m^2_{12}$ is given by
\begin{eqnarray}
m^2_{12}=\frac{v^2}{2|\delta|^2}\left\{
         {\cal I}(\delta\delta'){\cal I}(\tilde{\lambda}_6)
        +{\cal R}(\delta\delta')\sqrt{|\delta|^2-{\cal I}^2(\tilde{\lambda}_6)}        -|\delta|^2{\cal R}(\tilde{\lambda}_6)\right\}\,.\nonumber\\
\end{eqnarray}
We note that the induced phase vanishes if ${\rm arg}(A\mu)$, 
the sum of the phases $\Phi_\mu$ and $\Phi_{A}$, vanishes, even if
each of the CP--violating phases might not have to vanish. 
On the other hand, the size of $\delta$ or $\delta'$ is proportional to  
the pseudoscalar mass $m_a$ to a very good approximation so that if 
it becomes 
large, i.e, decoupled, the induced phase $\xi$ is diminished. Since the 
size of the induced phase is also inversely proportional to $\sin 2\beta$, 
the phase grows with increasing $\tan\beta$. In general, this induced phase 
will remain as a non--trivial physical 
phase and lead to a modification in the chargino and neutralino mass 
matrices.

Analytically, the induced phase $\xi$ palys a role of rotating the vacuum 
expectation value $v_2$ into $v_2\, {\rm e}^{i\xi}$. So, the chargino mass
matrix is given in the $(\tilde{W}^-,\tilde{H}^-)$ basis by
\begin{eqnarray}
{\cal M}_C=\left(\begin{array}{cc}
         M_2   &  -\sqrt{2}m_W\, c_\beta \\
         \sqrt{2}m_W\, s_\beta\, {\rm e}^{i\xi} & |\mu|\,{\rm e}^{i\Phi_\mu}
                 \end{array}\right)\,,
\end{eqnarray}
which is built up by the fundamental SUSY parameters; the SU(2) gaugino mass
$M_2$, the higgsino mass parameter $|\mu|$, its phase $\Phi_\mu$ and the
ratio $\tan\beta$ of the vacuum expectation values of the two neutral
Higgs doublet fields\footnote{We note that the vacuum expectation value of
the Higgs doublet $H_1$ has an opposite sign to the conventional one in
the literature. This is the reason why there appears a negative sign in 
the $(12)$ component of the chargino mass matrix.} 
The gaugino mass $M_2$ was made real already
by appropriate field redefinitions. Since the chargino mass matrix
${\cal M}_C$ is not symmetric, two different unitary matrices acting on
the left-- and right--chiral $(\tilde{W}^-,\tilde{H}^-)$ states are needed
to diagonalize the  matrix:
\begin{eqnarray}
U_{L,R}\left(\begin{array}{c}
             \tilde{W}^-  \\
             \tilde{H}^-
             \end{array}\right)
   =
       \left(\begin{array}{c}
             \tilde{\chi}^-_1 \\
             \tilde{\chi}^-_2
             \end{array}\right)\,,
\end{eqnarray}
and the mass eigenvalues are given by
\begin{eqnarray}
m^2_{\tilde{\chi}^\pm_{1,2}}=\frac{1}{2}\left[M^2_2+|\mu|^2+2m^2_W
     \mp \Delta\right]\,,
\end{eqnarray}
with $\Delta$ involving the phases $\Phi_\mu$ and the induced phase
$\xi$:
\begin{eqnarray}
\Delta=\left\{(M^2_2-|\mu|^2)^2+4m^4_W\cos^2 2\beta
              +4m^2_W(M^2_2+|\mu|^2-M_2|\mu|\sin 2\beta\cos(\Phi_\mu-\xi))
         \right\}^{1/2}\,.
\end{eqnarray}
As a matter of fact, an additional field redefinition enables
one to find that every CP--violating phenomenon due to the chargino 
mixing is dependent on the difference $\Phi_\mu-\xi$ of two 
phases $\Phi_\mu$ and $\xi$. Keeping in mind this point, one can 
infer the following aspects:
\begin{itemize}
\item Even if $\Phi_\mu$ vanishes there is still a source of CP violation
      due to the presence of the induced phase $\xi$ stemming from 
      the CP--violating phases in the scalar top and bottom sectors.
\item If both $\Phi_\mu$ and $\Phi_A$ vanish as in the CP--invariant
      theory, then the induced phase $\xi$ vanishes, leaving no source 
      of CP violation.
\end{itemize}
These effects may be investigated through the chargino pair production
as soon as the collider energies become high enough to go over their 
production thresholds. However, this investigation is beyond the regime
of the present work and so, it will be not touched upon here.

Similarly, the neutralino mass matrix describing neutralino mixing is 
modified by the introduction of the 
CP--violating induced phase. Following the same prescription for the
chargino mass matrix yields the neutralino mass matrix in the
$({\tilde B}, {\tilde W}^3, {{\tilde H}^0}_1,{{\tilde H}^0}_2 )$ basis:
\begin{eqnarray}
{\cal M}_N= \left(\begin{array}{cccc}
         |M_1|\, {\rm e}^{i\Phi_1} &    0             & 
          m_Z s_W c_\beta          &  m_Z s_W s_\beta\, {\rm e}^{i\xi} \\
          0                        &   M_2            & 
         -m_Z c_W c_\beta          & -m_Z c_W s_\beta\, {\rm e}^{i\xi} \\ 
          m_Z s_W c_\beta          & -m_Z c_W c_\beta & 
          0                        & -|\mu|\,{\rm e}^{i\Phi_\mu}        \\
          m_Z s_W s_\beta\,{\rm e}^{i\xi}  &
         -m_Z s_W s_\beta\,{\rm e}^{i\xi}  &
         -|\mu|\,{\rm e}^{i\Phi_\mu}       & 0
                   \end{array} \right)\,. 
\end{eqnarray}
The neutralino mass matrix ${\cal M}_N$ is a complex but symmetric
matrix so that it can be diagonalized by just one unitary matrix $N$
such that $N^*{\cal M}_N N^\dagger={\rm diag}(m_{\tilde{\chi}^0_1},
m_{\tilde{\chi}^0_2}, m_{\tilde{\chi}^0_3},m_{\tilde{\chi}^0_4})$
with the increasing ordering in masses. A simple orthogonality
transformation of a phase matrix enables one to confirm that
any physical observable related with the neutralino mixing
depends only on the phase $\Phi_1$ and the combination $\Phi_\mu-\xi$ 
of the phases $\Phi_\mu$ and $\xi$ as in the chargino system.
So, it is clear that except for the phase $\Phi_1$ the neutralino system
exhibits a similar dependence on the phase $\Phi_\mu$ and the induced
phase $\xi$. In the present work, we take the assumption of gaugino mass
unification for which the phase $\Phi_1$ should be zero at least up to
one--loop level against the renormalization group running from the
unification scale to the electroweak scale. In this scenario, there
exists only one CP--violating rephasing--invariant phase 
$\Phi_\mu-\xi$ in the chargino and neutralino mass sectors.

\section{Neutralino--nucleus elastic scattering}

\subsection{Four--Fermi effective Lagrangian}

In this section we investigate the importance of the CP-violating phases 
on the elastic scattering cross-sections of neutralinos on nuclei.
To this end, we calculate the four-Fermi effective $\chi$-quark interaction
Lagrangian with the inclusion of the CP violating phases
$\Phi_\mu$, $\Phi_1$, and $\xi$ for the standard spin--dependent and 
spin--independent neutralino--nucleus interactions. 
There are two types of diagrams contributing
to the elastic scattering; $Z$-exchange diagram and neutral Higgs boson
exchange diagrams as shown in Fig.~1. In order to derive the analytic 
expression for the scattering cross sections, it is necessary to determine 
the interactions of the $Z$ and  Higgs bosons to neutralinos and
fermions.

Firstly, the interactions of the neutral Higgs fields with SM 
fermions are described by the Lagrangian
\begin{eqnarray}
{\cal L}_{H\bar{f}f}=-H_{4-\alpha}\left\{
    \frac{gm_d}{2m_W s_\beta}
    \bar{d}\left[O_{2\alpha}-i s_\beta O_{1\alpha}\gamma_5\right]d
   +\frac{gm_u}{2m_W c_\beta}
    \bar{u}\left[O_{3\alpha}-i c_\beta O_{1\alpha}\gamma_5\right]u
    \right\}\,.
\end{eqnarray}
Here, $u$ denotes one of up--type fermions and $d$ one of
down--type fermions.  Obviously, the Higgs--fermion--fermion 
couplings are significant for the third--generation fermions, 
$t$, $b$ and $\tau$ because of their relatively
large Yukawa couplings. On the contrary, because any ordinary nucleus is 
mainly composed of the first (and second) generation fermions, the
couplings are very small. Nevertheless, these contributions become 
important for a nucleus with a large atomic/mass number, because they 
can contribute to the spin--independent cross section in a coherent manner.
We note in passing that the effect of CP--violating Higgs mixing 
is to induce a 
simultaneous coupling of $H_i$ ($i=1,2,3$) to CP--even and CP--odd 
fermion bilinears $\bar{f}f$ and $\bar{f} i\gamma_5 f$ \cite{DM}. 
This can lead to a sizable phenomenon of CP violation in the Higgs decays 
into polarized top-quark or tau-lepton pairs \cite{CKCK,CL}.

Secondly, the interactions of the Higgs bosons to neutralinos
involve the neutralino mixing and the Higgs boson mixing simultaneously.
As a result, the expression can become lengthy. So, for a notational 
convenience, 
we introduce the expression $G_\alpha$ defined in terms of the induced 
phase $\xi$, the neutralino diagonalization matrix $N$, and the Higgs 
diagonalization matrix $O$ as follows;
\begin{eqnarray}
G_\alpha=(N_{12}-t_W N_{11})\left[i(N_{13} s_\beta+N_{14} c_\beta\, 
         {\rm e}^{i\xi})O_{1\alpha}+N_{13} O_{2\alpha}
         +N_{14}\,{\rm e}^{i\xi} O_{3\alpha}\right]\,.
\end{eqnarray}
Then the interaction Lagrangian for the couplings of the neutral Higgs 
bosons to a lightest neutralino--pair is cast into a very simple form
\begin{eqnarray}
{\cal L}_{H\chi\chi}=\frac{g}{2}\sum_{\alpha=1}^3
       \bar{\chi}\bigg[{\cal R}(G_\alpha)+i {\cal I}(G_\alpha)\gamma_5
                 \bigg]\chi H_{4-\alpha}\,.
\end{eqnarray}
Here and from now on, we use  a simplified notation $\chi$ 
instead of the conventional notation $\tilde{\chi}^0_1$ to denote 
the lightest neutralino state.
We note that a sizable coupling is expected when the LSP has the 
significant compositions of both the gaugino states and the higgsino 
states. 

Thirdly, since both $\tilde{W}^0$ and $\tilde{B}$ have $T_3=Q=0$, these 
neutral gaugino states do not couple to the $Z$ boson. 
Therefore, the neutral current coupling of the $Z$ boson to neutralinos
occurs only from the higgsino component of neutralinos. 
The interaction Lagrangian is given by
\begin{eqnarray}
{\cal L}_{Z\chi\chi}=\frac{g}{4 c_W}\left[|N_{13}|^2-|N_{14}|^2\right]
                     \bar{\chi}\gamma^\mu\gamma_5\,\chi Z_\mu\,,
\end{eqnarray}
Note that the $Z$ boson couples to only an axial--vector current, 
reflecting the Majorana property of neutralinos.
One the other hand, the coupling of the $Z$ boson to fermions are
not changed even in the presence of new CP--violating phases.

Since the recoil momenta of the nucleus in the elastic scattering of the
neutralinos with fixed nuclei are very small
compared to the masses of the exchanged $Z$ and neutral Higgs bosons,
it is appropriate to have an effective four--Fermi Lagrangian 
by taking the momentum transfer to be zero. The general form of
the four-Fermi effective Lagrangian can be written as
\begin{eqnarray}
{\cal L}&=&\alpha_{1f}\left(\bar{\chi}\gamma^\mu\gamma_5 \chi\right) 
                      \left(\bar{f}\gamma_\mu f\right)
          +\alpha_{2f}\left(\bar{\chi}\gamma^\mu\gamma_5 \chi\right) 
                      \left(\bar{f}\gamma_\mu\gamma_5 f\right)
          +\alpha_{3f}\left(\bar{\chi}\chi\right) 
                      \left(\bar{f}f\right) \nonumber\\
        &&+\alpha_{4f}\left(\bar{\chi}\gamma_5\chi\right) 
                      \left(\bar{f}f\right)
          +\alpha_{5f}\left(\bar{\chi}\chi\right) 
                      \left(\bar{f}\gamma_5f\right) 
          +\alpha_{6f}\left(\bar{\chi}\gamma_5\chi\right) 
                      \left(\bar{f}f\right)\,.
\end{eqnarray}
The effective Lagrangian should be summed over fermions and the 
coefficients $\alpha_{if}$ ($i=1$ to 6) based on the
effective Lagrangian can be obtained by evaluating two Feynman diagramsi
in Fig.~1 as 
\begin{eqnarray}
&&\alpha_{1f}=+\frac{G_F}{\sqrt{2}}\left[|N_{13}|^2-|N_{14}|^2\right]
               (T^f_3-2Q_f s^2_W)\,,\nonumber\\
&&\alpha_{2f}=-\frac{G_F}{\sqrt{2}}\left[|N_{13}|^2-|N_{14}|^2\right]
              T^f_3\,,\nonumber\\
&&\alpha_{3f}=-\frac{gY_f}{2\sqrt{2}}\sum_{\alpha=1}^{3}
               \frac{{\cal R}(G_\alpha)}{m^2_{H_{4-\alpha}}}
               \left\{\begin{array}{cl}
                      O_{3\alpha} & {\rm for}\ \ u \\
                      O_{2\alpha} & {\rm for}\ \ d
                      \end{array}\right.\,,\nonumber\\
&&\alpha_{4f}=-\frac{gY_f}{2\sqrt{2}}\sum_{\alpha=1}^{3}
               \frac{{\cal I}(G_\alpha)}{m^2_{H_{4-\alpha}}}
               O_{1\alpha}\left\{\begin{array}{cl}
                                  c_\beta & {\rm for}\ \ u \\
                                  s_\beta & {\rm for}\ \ d
                                  \end{array}\right.\,,\nonumber\\
&&\alpha_{5f}=+\frac{igY_f}{2\sqrt{2}}\sum_{\alpha=1}^{3}
               \frac{{\cal R}(G_\alpha)}{m^2_{H_{4-\alpha}}}
               O_{1\alpha}\left\{\begin{array}{cl}
                                  c_\beta & {\rm for}\ \ u \\
                                  s_\beta & {\rm for}\ \ d
                                  \end{array}\right.\,,\nonumber\\
&&\alpha_{6f}=-\frac{igY_f}{2\sqrt{2}}\sum_{\alpha=1}^{3}
               \frac{{\cal I}(G_\alpha)}{m^2_{H_{4-\alpha}}}
               \left\{\begin{array}{cl}
                      O_{3\alpha} & {\rm for}\, u \\
                      O_{2\alpha} & {\rm for}\, d
                      \end{array}\right.\,.
\end{eqnarray}
In these expressions, $G_F$ is the Fermi constant, $Y_f$ the Yukawa coupling
of the fermion $f$, which is $gm_u/(\sqrt{2}m_W s_\beta)$ for $u$--type
fermions and $gm_d/(\sqrt{2}m_W c_\beta)$ for $d$--type fermions,
and $T^f_3$ the third component of the isospin of the fermion $f$. 
In the limit of vanishing CP-violating phases, these expressions
agree with those in \cite{JKG} and \cite{EF}.   

Among the six independent terms in the effective Lagrangian, only the 
terms with the coefficients $\alpha_{2f}$ and $\alpha_{3f}$ survive 
in the vanishing momentum--transfer limit. The coefficient $\alpha_{2f}$ 
contains contributions from the $Z$ boson exchange while the coefficient 
$\alpha_{3f}$ has contributions from the neutral Higgs--boson exchanges. 
The spin--dependent contribution from the $\alpha_{2f}$ terms 
contains are not suppressed by the fermion mass and can be large
over much of the parameter space.
In contrast, the spin--independent contribution from the $\alpha_{3f}$ terms 
is always proportional to fermion masses and relies on the LSP being 
a well--balanced mixture of gaugino and higgsino states and the size
of the scalar--pseudoscalar mixing. It might be naively expected that
the small first and second generation fermion masses will give rise
to a very small spin--independent cross section. However, the 
spin--independent cross-section can be enhanced by the effects of 
coherent scattering in a nucleus and can dominate over the 
spin--dependent cross-section for heavy nuclei. In the following subject, 
we present the analytic form of both the spin--dependent and 
spin--independent elastic cross sections and investigate the 
physical parameters determining them.

\subsection{Elastic cross sections: spin--dependent 
            versus spin--independent}

The elastic scattering cross sections based on $\alpha_{2,3f}$ have been
conveniently expressed in \cite{JKG}.  The spin--dependent cross-section
can be written as
\begin{eqnarray}
\sigma_{_S} = \frac{32}{\pi} G_F^2 m_r^2 \Lambda^2 J (J+1)\,,
\end{eqnarray}
where $m_r=m_fm_N/(m_f+m_N)$ is the reduced neutralino-nucleus mass, 
$J$ is the spin of the nucleus, the value of which is $4.5$ for $^{73}$Ge 
and $0.5$ for $^{19}$F, respectively, and the quantity $\Lambda$ is given by 
\begin{eqnarray}
\Lambda=\frac{1}{J}\left(a_p\langle S_p\rangle
                         +a_n\langle S_n\rangle\right)\,,
\end{eqnarray}
with the coefficients $a_p$ and $a_n$: 
\begin{eqnarray}
a_p = \sum_{f=u,d,s}\frac{\alpha_{2f}}{\sqrt{2}G_F} \Delta^{(p)}_f, \qquad 
a_n = \sum_{f=u,d,s} \frac{\alpha_{2f}}{\sqrt{2}G_F} \Delta^{(n)}_f\,.
\label{as}
\end{eqnarray}
The factors $\Delta^{(p,n)}_f$ depend on the spin content of the nucleus 
and the values of the factors are taken to be \cite{SMC} 
\begin{eqnarray}
&&\Delta^{(p)}_u =+0.77\,,\qquad
  \Delta^{(p)}_d =-0.38\,,\qquad
  \Delta^{(p)}_s =-0.09\,, \nonumber\\
&&\Delta^{(n)}_u =-0.38\,,\qquad
  \Delta^{(n)}_d =+0.77\,,\qquad
  \Delta^{(n)}_s =-0.09\,, 
\end{eqnarray}
in our analysis. 
The expectation values $\langle S_{p,n}\rangle$ are the averaged values 
of the spin content 
in the nucleus and therefore are dependent on each target nucleus.  
We will display results for scattering off of a $^{73}$Ge target and
a $^{19}$F for which in the shell model 
\begin{eqnarray}
&& \langle S_{p}\rangle_{\rm Ge} = 0.011\,,\qquad
   \langle S_{n}\rangle_{\rm Ge} = 0.491\,,\nonumber\\
&& \langle S_{p}\rangle_{\rm F}\,\,= 0.415\,,\qquad 
   \langle S_{n}\rangle_{\rm F}\,=-0.047\,. 
\end{eqnarray}
For a more detailed information on the these quantities, we refer to the
review paper by Jungman, Kamionkowski and Griest \cite{JKG}. 

On the other hand, the spin--independent cross section is written as
\begin{eqnarray}
\sigma_{_I}=\frac{4m_r^2}{\pi}\left[Zf_p+(A-Z)f_n\right]^2\,,
\end{eqnarray}
where $Z$ and $A$ are the atomic number and the mass number of the nucleus,
respectively, and the coefficients $f_p$ are given by
\begin{eqnarray}
\frac{f_p}{m_p}=\sum_{q=u,d,s} f_{Tq}^{(p)} \frac{{\alpha_3}_q}{m_q} 
               +\frac{2}{27}f_{TG}^{(p)} \sum_{q=c,b,t} 
                \frac{{\alpha_3}_q}{m_q}\,,
\label{fp}
\end{eqnarray}
and $f_n$ is given by an expression similar to that for $f_n$. 
The parameters 
$f_{Tq}^{(p)}$ are defined by $\langle p|m_q{\bar q}q|p\rangle 
= m_p f_{Tq}^{(p)}$, while $f_{TG}=1-(f_{Tu}+f_{Td}+f_{Ts})$ \cite{SVZ}.  
For our numerical analysis, we adopt \cite{GLS}
\begin{eqnarray}
&& f_{Tu}^{(p)} = 0.019\,,\qquad
   f_{Td}^{(p)} = 0.041\,,\qquad
   f_{Ts}^{(p)} = 0.140\,, \nonumber\\
&& f_{Tu}^{(n)} = 0.023\,,\qquad
   f_{Td}^{(n)} = 0.034\,,\qquad
   f_{Ts}^{(n)} = 0.140\,.
\end{eqnarray}
There exist additional contributions due to one--loop Higgs couplings 
to gluons and so-called twist-2 operators; however the change from a more 
careful treatment of loop effects for heavy
quarks and the inclusion of twist-2 operators is expected to be
numerically small \cite{DN1,DN2}.

\section{LSP--nucleus elastic scattering cross sections}

\subsection{Independent SUSY parameters}

The elastic scattering cross sections depends on a large number of
SUSY parameters; more than ten parameters. So, in order to make
a realistic analysis, it will be necessary to make a few assumptions
which are reasonable for physics point of view. 

Firstly, we note that the neutral Higgs mass spectrum depends on 
the chargino mass $m_{H^\pm}$, a SUSY breaking scale $M_{\rm SUSY}$, 
two trilinear terms 
$|A_{t,b}|$ and their phases $\Phi_{A_t,A_b}$ as well as 
$\tan\beta$, the higgsino mass parameter $|\mu|$ and its phase
$\Phi_\mu$. Let us assume a universal trilinear parameter in 
our analysis:
\begin{eqnarray}
|A_t|=|A_b|\equiv |A|\,, \qquad \Phi_{A_t}=\Phi_{A_b}\equiv \Phi_A\,,
\end{eqnarray}
This assumption will not forbid us from finding out the general trend
of the Higgs masses and their couplings to fermions and neutralinos.
The size of the Higgs boson mixing is determined by the dimensionless 
parameter $\eta_{_{CP}}$. We take in our analysis 
\begin{eqnarray}
M_{\rm SUSY}=0.5\,{\rm TeV}\,,\qquad
|A|=1.5\,{\rm TeV}\,.
\end{eqnarray}
The charged Higgs mass plays a crucial role
in determining the contribution of the spin--independent cross
section so that it will be treated as a free parameter along with
the parameters $\{\tan\beta, |\mu|,\Phi_\mu\}$.

Secondly, the neutralino masses and mixing are determined by the
SU(2) gaugino mass parameter $M_2$, the U(1) gaugino mass $|M_1|$ and
its phase $\Phi_1$ as well as the parameters $\{\tan\beta,|\mu|,\Phi_\mu\}$.
It will be reasonable to take the gaugino mass unification
condition between two gaugino masses so that at least up to the one--loop
level one have
\begin{eqnarray}
|M_1|=\frac{5}{3}t^2_W M_2\sim 0.5 M_2\,,\qquad \Phi_1=0\,,
\end{eqnarray}
where $t_W=\tan\theta_W$.
In this case, the lightest neutralino will be Bino--like for
$|\mu|\gg M_2$ and it will be higgsino--like for $|\mu|\ll M_1$.
Note that the higgsino parameter $|\mu|$ and its phase $\Phi_\mu$
affect both the neutralino mixing and neutral Higgs mixing.
For a large Higgs mixing, a large $|\mu|$ is preferred. On the contrary,
a large neutralino mixing requires a relatively small value of $|\mu|$.
In this light, $|\mu|$ is a crucial SUSY parameter in determining
the relative importance of the spin--dependent and spin--independent
contributions. We will take $|\mu|$ and $\Phi_\mu$ as free parameters.

Thirdly, the spin--independent cross section strongly depends on 
fermion masses. For the fermion masses,
we will use their maximum values coded by the Particle Data Group
\cite{PDG}:
\begin{eqnarray}
&& m_u=5.0\,{\rm MeV}\,,\qquad\!\! m_d=9.0\,{\rm MeV}\,,\qquad
   m_s= 170\,{\rm MeV}\,,\nonumber\\
&& m_c=1.4\,{\rm GeV}\,,\qquad m_b=4.4\,{\rm GeV}\,,\qquad 
   m_t =174\,{\rm GeV}\,.
\end{eqnarray}
It is certain that there still exist large uncertainties in the values
of the fermion masses. Nevertheless,
our numerical analysis will be qualitatively reasonable
and  even quantitatively meaningful with some improved determinations
of fermion masses.

Consequently, the independent SUSY parameters which we manipulate
in our numerical estimates of the LSP--nucleus scattering cross
sections are 
\begin{eqnarray}
\tan\beta\,,\qquad M_2\,,\qquad |\mu|\,,\qquad m_{H^\pm}\,,\qquad 
\Phi_\mu\,, \qquad \Phi_A\,.
\end{eqnarray}
We set $\Phi_\mu=0$ and take two values 3 and 30 for $\tan\beta$
which will satisfy the constraints from the
Higgs search experiments at LEP\cite{HIGGS} and the 2--loop Barr--Zee--type
electron and neutron EDMs.

\subsection{Numerical results} 

We are now ready to show the importance of the CP--violating phases
in the LSP detection through the LSP--nucleus elastic scattering.
For a systematic analysis, it is useful to understand the
dependence of the CP--violating induced phase on the parameters
$|\mu|$, $\Phi_\mu$ and $\Phi_A$. Incidentally, the induced phase
depends only on one combination of two phases $\Phi_\mu+\Phi_A$.
So, for this analysis, we simply introduce $\Phi=\Phi_\mu+\Phi_A$
and set the phase to be $\pi/2$, which will give (almost) the \
maximal absolute value of $\sin\xi$. In this case, the sine of the
induced phase is determined by the higgsino mass parameter $|\mu|$ 
and the charged Higgs mass for which we take into account two
values; 250 GeV and 500 GeV. Also, $\sin\xi$ in this limit is
is given by a simple analytical expression
\begin{eqnarray}
\sin\xi=-\frac{|\tilde{\lambda}_6|}{m^2_a\sin 2\beta}\,.
\end{eqnarray}
Therefore, $\sin\xi$ is always negative and it is expected that
the existence of $\sin 2\beta$ in the denominator forces its absolute 
value to increase with increasing $\tan\beta$.

Figure~1 shows $|\sin\xi|$ as a function of the higgsino mass parameter
$|\mu|$ by taking four different sets of $\{\tan\beta, m_{H^\pm}\}$; 
$\{3, 250\,{\rm GeV}\}$ (solid line),
$\{30,250\,{\rm GeV}\}$ (dashed line), $\{3,500\,{\rm GeV}\}$ 
(dot--dashed line), and $\{30,500\,{\rm GeV}\}$ (dotted line).
The values of the other SUSY parameters are given in the previous
section. It is clear that $|\sin\xi|$ increases with increasing
$\tan\beta$ but decreases with increasing the charged Higgs mass
$m_{H^\pm}$. It depends very strongly on the higgsino mass parameter
$|\mu|$; for a small $\tan\beta$ the absolute value of $\sin\xi$ 
can be at most as large as 0.1 for a large value of $|\mu|$ and
for a relatively small value of $m_{H^\pm}$, but it can be
as large as unity for a large value of $|\mu|$ and a small
value of $m_{H^\pm}$. This is due to the fact that for a small
value of $\tan\beta$ the scalar top quarks contribute (almost) 
exclusively, but for a large value of $\tan\beta$ the scalar bottom
quarks as well as the scalar top quarks contribute to the CP--violating
induced phase. In this light, one may expect that the CP--violating induced 
mass $\xi$ can affect the chargino and neutralino sectors significantly 
for a certain regime of the SUSY parameter space. However, we note that
the relative size of the induced phase contribution to the chargino 
or neutralino mass spectrum is at most as large as
$(m_W/|\mu|)^2\sin 2\beta$ for $|\mu|$ much larger than $M_2$
so that the enhancement effect in the induced phase by a large $|\mu|$ and
a large $\tan\beta$ is washed out through the strong suppression 
by the same parameters. Therefore, in most cases there are no significant 
modifications in the chargino and neutralino mass spectrums.

To begin with, we estimate the spin--dependent cross section. 
For this case we consider the scattering of neutralinos on fluorine
$^{19}F$, for which the spin--dependent contribution typically dominates
by a factor of 20 \cite{JKG}. As shown in Section~2B, the crucial parameters
determining the spin--dependent cross section are the coefficients 
$\alpha_{2f}$ and on the whole by the coefficients $a_p$ and $a_n$.
Because there is simply one $Z$--exchange contribution to the
spin--independent process, all the $\alpha_{2f}$ coefficients are
proportional to the coupling of the $Z$ boson to neutralinos
except for their relative factors; $|N_{13}|^2-|N_{14}|^2$. 
We find by a comprehensive numerical scan on the relevant
SUSY parameters that the CP--violating induced phase {\it hardly} changes
the values of $a_p$ and $a_n$. This property is in a sharp
contrast with the aspect that the coefficients and the spin--dependent
cross section are very sensitive to $\Phi_\mu$, the phase of the 
higgsino. For the details of the significant dependence on the
phase $\Phi_\mu$, we refer to
the work by Falk, Ferstl and Olive \cite{FO}.

On the other hand, the spin--independent cross section is dominant 
in much of the SUSY parameter space for scattering off heavy nuclei
so that we consider the scattering of neutralinos on
$^{73}$Ge. The cross section is determined by the coefficients
$\alpha_{3f}$ which involve the contributions from the
neutral Higgs exchanges and the scalar part of their couplings
to neutralinos and fermions. Because the explicit CP violation
through radiative corrections to the Higgs sector can lead to a large
mixing among scalar Higgs bosons and pseudoscalar Higgs boson,
it is naturally expected to exhibit a rather strong dependence of the
cross section on the phases such as the phase $\Phi_A$ and the
induced phase $\xi$, even if the phase $\Phi_\mu$ is taken to be
zero in favor of the naturalness conditions at the GUT scale.
Nevertheless, the structure of the coupling $G_\alpha$ requires
the lightest neutralino state to be composed of both gauginos
and higgsinos with significant compositions. This means that
a large cross section can be obtained when the gaugino mass $M_2$
and the higgsino mass parameter $|\mu|$ are comparable in size.

Keeping in mind these aspects, we take for our numerical analysis
two values of $\tan\beta$ (3 and 30), set $m_{H^\pm}=250\,{\rm GeV}$
and use the same values for the other SUSY parameters as
those given in the previous analyses. Figure~3 shows $f_p$\footnote{The
dependence of $f_n$ on the phase $\Phi$ and the parameter $|\mu|$ is very
similar to that of $|f_p|$ and furthermore we find that two quantities
have a very similar value for the whole scanned space of the SUSY
parameters.} for five values of the higgsino mass parameter $|\mu|$;
200 GeV (solid line), 400 GeV (long dashed line), 600 GeV (dot--dashed
line), 800 GeV (dotted line) and 1000 GeV (dashed line).
The value of $\tan\beta$ is taken to be 3 in the left frame and
30 in the right frame, respectively. We note several interesting aspects
from two figures:
\begin{itemize}
\item The magnitude of $f_p$ (as well as $f_n$) is very sensitive to
      the value of $|\mu|$. In most cases except for the region
      of large values of $\Phi$, it decreases with increasing
      $|\mu|$. The main reason for the suppression is that the LSP
      is gaugino--dominated for a large value $|\mu|$, while 
      an intermediate state of the LSP is required to have a sizable
      $f_p$.
\item Comparing the small and large $\tan\beta$ cases, one can find that
      the magnitude of $f_p$ increases with increasing $\tan\beta$. 
\item The relative sensitivity of $f_p$ to the phase $\Phi$
      is very much enhanced for a large value of $|\mu|$, in particular
      for a small value of $\tan\beta$. In this case, $f_p$
      increases by one order of magnitude as $\Phi$ changes from zero to
      180 degrees. As a result, one can expect to have a large LSP
      detection rate even for a large $|\mu|$.
\item On the contrary, for a large $\tan\beta$ and a large $|\mu|$,
      the value of $f_p$ tends to be suppressed for non--trivial values
      around $\Phi=90^0$.
\end{itemize}
On the whole, it is clear that $f_p$ is very sensitive to $|\mu|$ and
$\tan\beta$, and it becomes sensitive to the phase $\Phi$, equivalently,
$\Phi_A$ in the present work, for a large $|\mu|$. 
However, the behavior of $f_p$ with
respect to the phase $\Phi$ strongly depends on the value of $\tan\beta$.

From the discussion in the previous paragraph, it is expected that the
spin--independent cross section also will show the same behavior with
respect to the SUSY parameters and CP--violating phases. 
With the fact that $f_p$ and $f_n$ are (almost) similar to each other
in size, one can see that the spin--independent cross section is
proportional to the square of $f_p$ so that the dependence of the
cross section is enhanced. Figure~4 shows the dependence of the
spin--independent cross section $\sigma_I(\chi {}^{73}{\rm Ge}\rightarrow
\chi {}^{73}{\rm Ge})$ on the phase $\Phi$ with the other SUSY parameters as
those in Figure~3. As expected, the behavior of the cross section 
is similar to that of $f_p$. For $\tan\beta=3$ and $|\mu|=1000$ GeV,
the cross section changes by one order of magnitude, depending on the
phase, while the cross section tends to be suppressed for non--trivial
phases in the case of a large $\tan\beta$.

\section{Summary and Conclusions}

In this paper, we have re--visited the neutralino--nucleus elastic 
scattering process portion of dark matter in the Universe under a
specific SUSY scenario (which has been recently suggested to avoid the 
severe EDM constraints); in the scenario, the first and
second generation sfermions are so heavy that they are decoupled from
the low--energy supersymmetric theories, while the third generation 
sfermions are required to be relatively light not to spoil naturalness.
In this case, the complex parameters, in particular, the phase of
the trilinear terms and the phase of the higgsino mass parameter, of
the third--generation stop and sbottom sectors can lead to a significant
mixing between scalar neutral Higgs bosons and a pseudoscalar neutral 
Higgs boson.
As a result, there appears a CP--violating induced phase due to the
misalignment between two Higgs doublet fields. We have focussed on
the impact of the scalar--pseudoscalar mixing and the induced phase
on the LSP--nucleus elastic scattering process.  

For the sake of our numerical analysis, we have assumed a universal
trilinear parameters $|A|$ and $\Phi_A$ and set the phase of the higgsino
mass parameter $\Phi_\mu$ to be zero, and then we have varied the
charged Higgs boson mass, $\tan\beta$, $|\mu|$ and the phase $\Phi_A$ while
keeping $|A|=2$ TeV and $M_{\rm SUSY}=0.5$ TeV, the SUSY breaking scale. 
To summarize, we have several interesting aspects related with the 
induced phase and the detection of the neutralino dark matter:
\begin{itemize}
\item The induced phase $\xi$ itself can be large if $\tan\beta$ is
      large, and $|\mu|$ is comparable to $M_{\rm SUSY}$ or larger.
      However, we have found that in this case its contribution to 
      chargino and neutralino masses is strongly suppressed because 
      the contribution is determined by
      the combination $(m_W/|\mu|)^2\sin 2\beta$. 
\item The coupling of the $Z$ boson to neutralinos is affected only
      through the CP--violating induced phase from the third generation
      stop and sbottom sectors. We have found that because of the mutually
      destructive properties mentioned before the CP--violating induced 
      phase hardly changes the spin--dependent cross section. 
\item The spin--independent part, to which the main contribution comes
      from the neutral Higgs boson exchanges can undergo a big change
      with respect to the phase $\Phi_A$ and the induced phase $\xi$
      as well. The Higgs mass spectrum and the couplings of the neutral
      Higgs bosons to neutralinos and fermions vary very significantly
      with respect to the phases as well as the other SUSY parameters.
      The dimensionful parameter $f_p$ as well as $f_n$ dictating the 
      size of the spin--independent cross section increases with 
      decreasing $|\mu|$ and increasing $\tan\beta$, while the sensitivity 
      of the parameter to the phase $\Phi$. For a small value of 
      $\tan\beta=3$, the sensitivity of $f_p$ and the spin--independent 
      cross section to the phase $\Phi$ can be huge for a large value 
      of $|\mu|$. On the contrary, the spin--independent cross section
      is suppressed for non--trivial values of the phase $\Phi$
      for a large value of $\tan\beta$.
\end{itemize}
Even though we have not presented here, we have confirmed the result
by Falk, Ferstl and Olive \cite{FO} that the spin--dependent and 
spin--independent  cross sections are strongly dependent on the phase 
$\Phi_\mu$. In some cases, for a broad range of non--zero $\Phi_\mu$, there are
cancellations in the cross sections which reduce both the spin--dependent
and spin--independent cross sections by more than an order of magnitude.
In other cases, there may be enhancements as one varies $\Phi_\mu$.

In general there are several CP--violating phases which
can affect many important physics phenomena. In addition to the
supersymmetric CP--violating phase $\Phi_\mu$, we have found that
even in the scenario of the gaugino mass unification and the 
decoupling of the first and second sfermion states, the CP--violating
scalar--pseudoscalar neutral Higgs boson mixing and the CP--violating
induced phase between two Higgs doublets can affect the spin--independent
part of the neutralino--nucleus elastic scattering cross section
significantly, depending on the values of $\tan\beta$, the charged Higgs
boson mass, the SUSY breaking scale, and so on.
Therefore, we can conclude that in the present situation of no SUSY
signatures it is important to clearly understand the impact of all the 
CP--violating phases, (which are not constrained by low--energy 
measurements such as the electron and neutron EDMs), on the
neutralino--nucleus elastic scattering, one of the most promising
dark matter detection mechanism.

\vskip 0.7cm

\subsection*{Acknowledgments}

The author is grateful to Manuel Drees and Toby Falk for helpful 
discussions and thanks Francis Halzen and the Physics Department,
University of Wisconsin--Madison where the work was initiated.
The author wishes to acknowledge the financial support of 1997--sughak 
program of Korea Research Foundation. 

\vskip 3.5cm

\noindent
{\large\bf Note added}

\vskip 0.3cm
While finalizing our paper, we became aware of one very recent work \cite{GF}
by Gondolo and Freese which treats some of the features we have been studying 
here. In the paper, they have mainly concentrated on the scalar--pseudoscalar
mixing and have provided a global parameter scan in estimating the elastic 
scattering cross section. Instead, our work have studied the effect of 
the CP--violating induced phase between two Higgs doublets as well as 
of the scalar--pseudoscalar mixing.


%

%


\begin{figure}
 \begin{center}
\mbox{ }
\vskip 2cm
\hbox to\textwidth{\hss\epsfig{file=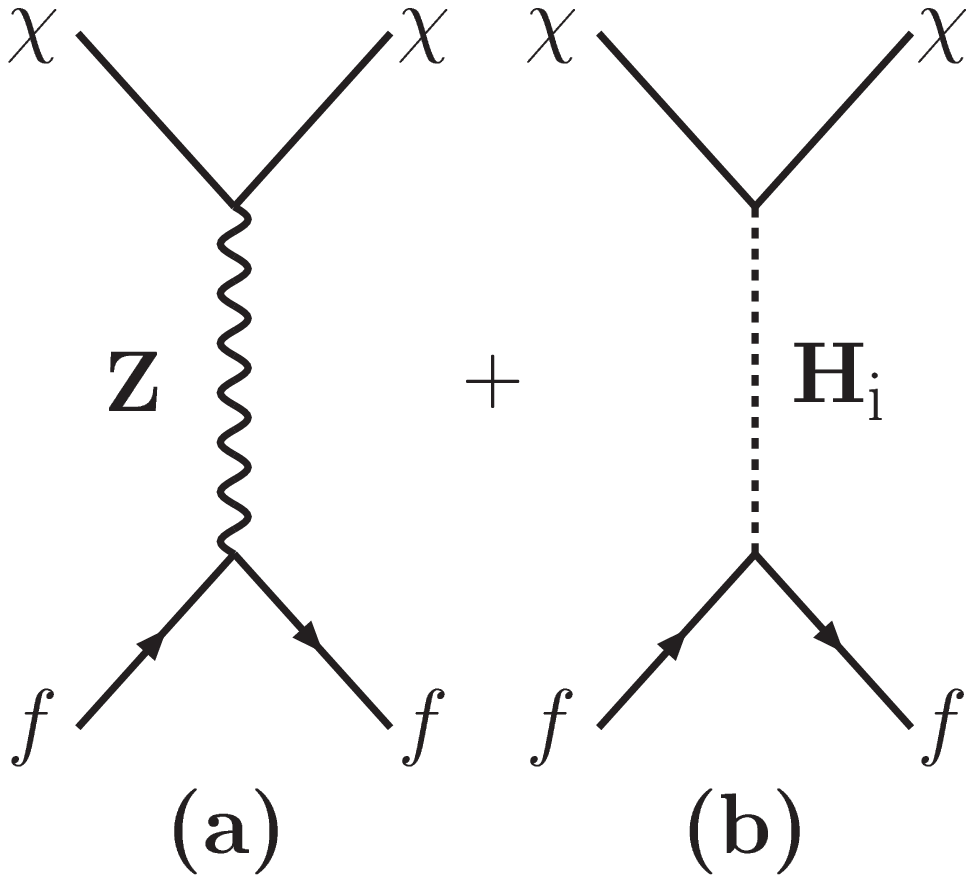,width=20cm,height=16cm}\hss}
 \end{center}
\vskip -6cm
\caption{Two types of Feynman diagrams contributing to the 
         neutralino--nucleus elastic scattering process; (a) 
         the spin--dependent $Z$-exchange diagram
         and (b) the spin--independent Higgs--boson--exchange diagram.}
\label{fig:diagram}
\end{figure}


\begin{figure}
 \begin{center}
\hbox to\textwidth{\hss\epsfig{file=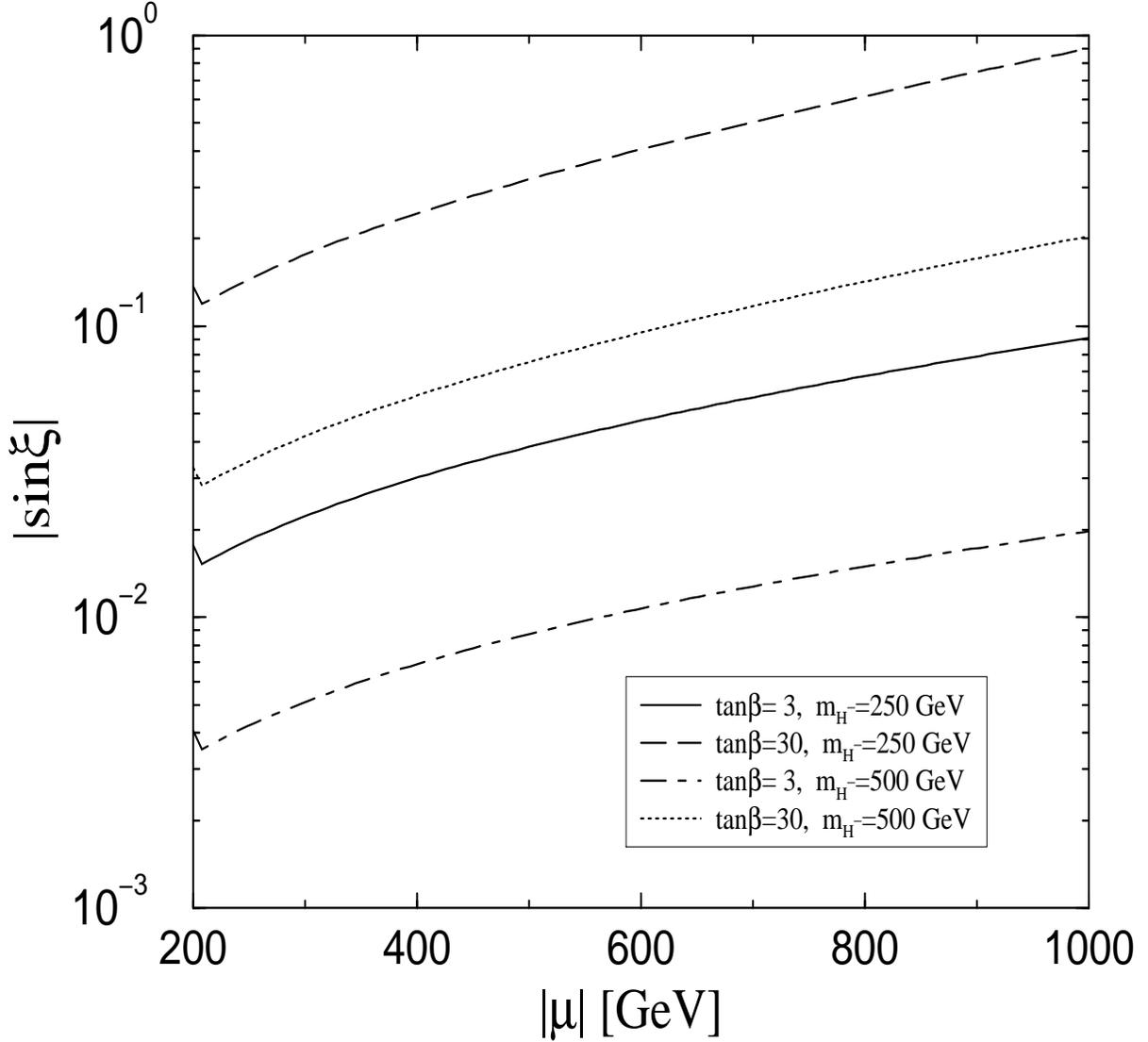,width=16cm,height=15cm}\hss}
 \end{center}
\caption{The absolute value $|\sin\xi|$ of the CP--violating induced 
         phase $\xi$ for $\Phi=\pi/2$ as a function
         of the higgsino mass parameter $|\mu|$. We take four different
         sets of $\{\tan\beta,m_{H^\pm}\}$; $\{3,250\,{\rm GeV}\}$ (solid
         line), $\{30,250\,{\rm GeV}\}$ (dashed line),
         $\{3,500\,{\rm GeV}\}$ (dot--dashed line) and
         $\{30,500\,{\rm GeV}\}$ (dotted line). The values of the other
         SUSY parameters are given in the text.}
\label{fig:xi}
\end{figure}


\begin{figure}
 \begin{center}
\hbox to\textwidth{\hss\epsfig{file=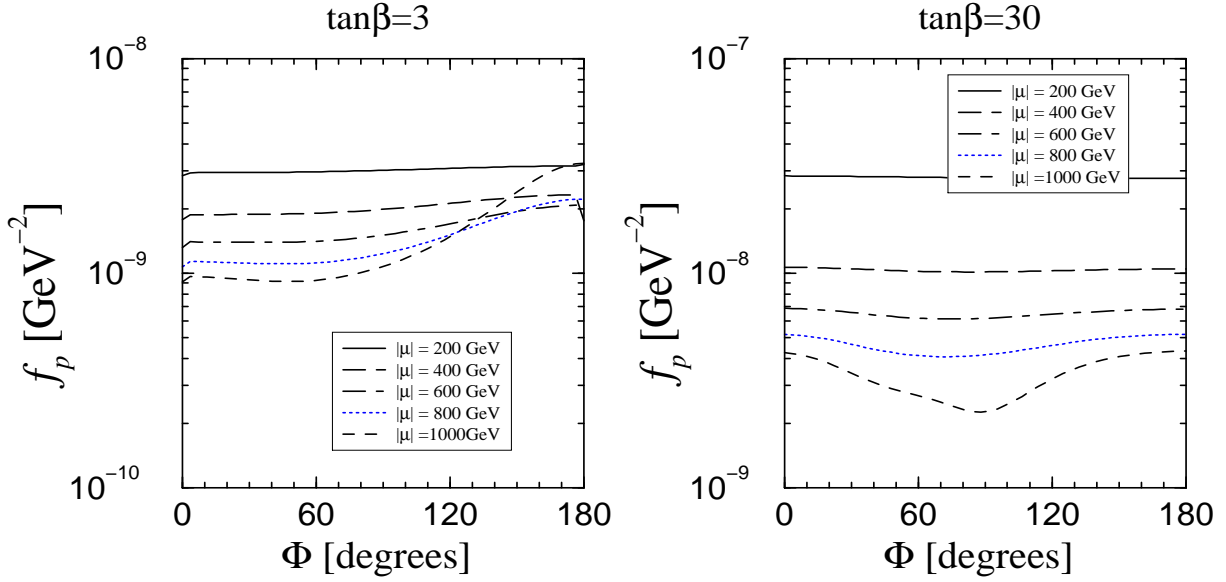,width=16cm,height=8cm}\hss}
 \end{center}
\caption{The dimensionful coefficient $f_p$ as a function of the phase 
         $\Phi$ with $\Phi_\mu=0$ for the 
         spin--independent neutralino--nucleus elastic scattering.
         The solid line is for $|\mu|=200$ GeV, the long dashed line
         for $|\mu|=400$ GeV, the dot--dashed line for $|\mu|=600$ GeV,
         the dotted line for $|\mu|=800$ GeV and the dashed line
         for $|\mu|=1000$ GeV. The value of $\tan\beta$ is taken
         to be 3 in the left frame and 30 in the right frame. In both
         cases, the charged Higgs boson mass is 250 GeV.
         The values of the other remaining SUSY parameters are given 
         in the text.}
\label{fig:sifp}
\end{figure}


\begin{figure}
 \begin{center}
\hbox to\textwidth{\hss\epsfig{file=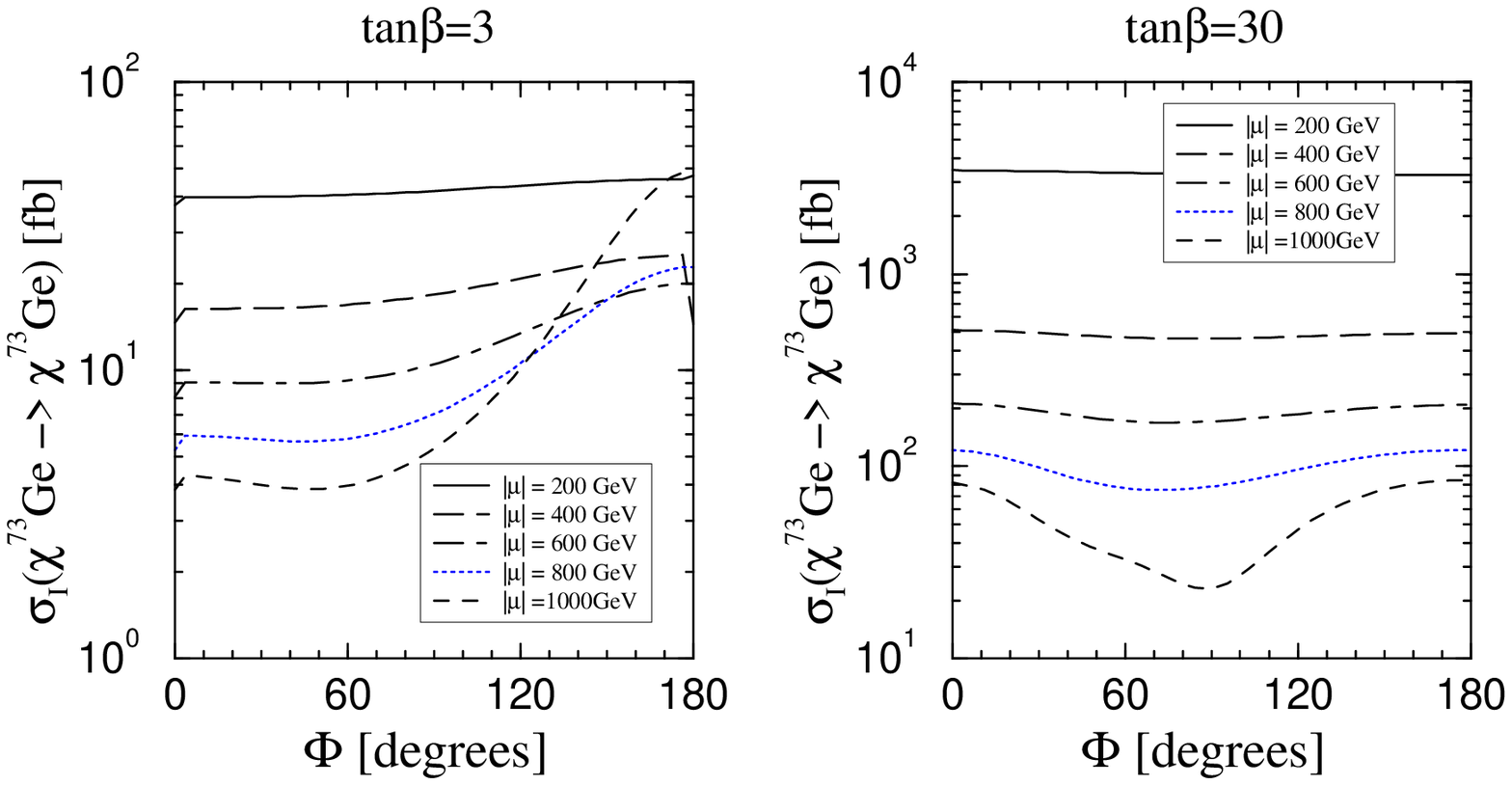,width=16cm,height=8cm}\hss}
 \end{center}
\caption{The spin--independent elastic scattering cross
         section $\sigma_{_I}(\chi {}^{73}{\rm Ge}\rightarrow 
         \chi {}^{73}{\rm Ge})$
         as a function of the phase $\Phi$ with $\Phi_\mu=0$ for the 
         spin--independent neutralino--nucleus elastic scattering.
         The legend for the lines is the same as that in Fig.~3.}
\label{fig:xsec}
\end{figure}

\vfil\eject


\begin{references}
\bibitem{MSSM} 
   For reviews, see H.~Nilles, Phys.~Rep.~{\bf 110}, 1 (1984); 
   H.E.~Haber and G.L.~Kane, Phys.~Rep.~{\bf 117}, 75 (1985);
   S.~Martin, in {\it Perspectives on Supersymmetry}, edited by G.L.~Kane,
   (World Scientific, Singapore, 1998). 
\bibitem{MS} A.~Masiero and L.~Silvetrini, in {\it Perspectives on 
   Supersymmetry}, edited by G.L.~Kane, (World Scientific, Singapore, 
   1998); J.~Ellis, S.~Ferrara and D.V.~Nanopoulos, 
   Phys.~Lett.~B{\bf 114}, 231 (1982); W.~Buchm\"{u}ller and D.~Wyler,
   {\it ibid.}~B{\bf 121}, 321 (1983); J.~Polchinski and M.B.~Wise,
   {\it ibid.}~B{\bf 125}, 393 (1983); F. del Aguila, M. Gavela, J. Grifols, 
   and A. Mendez, Phys. Lett. {\bf B126}, 71 (1983); 
   J.M.~Gerard {\it et al.}, Nucl.~Phys.~{\bf B253}, 93 (1985); 
   P.~Nath, Phys.~Rev.~Lett.~{\bf 66}, 2565 (1991); 
   R.~Garisto, Nucl.~Phys.~{\bf B419}, 279 (1994); D.V. Nanopoulos 
   and M. Srednicki, Phys. Lett. {\bf B128}, 61 (1983). 
\bibitem{eEDM} E.D.~Commins, S.B.~Ross, D.~DeMille, and B.S.~Regan, 
   Phys.~Rev.~A {\bf 50}, 2960 (1994); K.~Abdullah et al., 
   Phys.~Rev.~Lett.~{\bf 65}, 2347 (1990).
\bibitem{KO} Y.~Kizukuri and N.~Oshimo, Phys.~Rev.~D {\bf D45}, 1806 
   (1992); {\bf 46}, 3025 (1992).
\bibitem{Kaplan} S.~Dimopoulos and G.F.~Giudice, Phy.~Lett.~B {\bf 357}, 
   573 (1995); A.~Cohen, D.B.~Kaplan and A.E.~Nelson, {\it ibid.}~B 
   {\bf 388}, 599 (1996); A.~Pomarol and D.~Tommasini, 
   Nucl.~Phys.~{\bf B466}, 3 (1996).
\bibitem{IN} T.~Ibrahim and P.~Nath, Phys.~Rev.~D {\bf 57}, 478 (1998); 
   M.~Brhlik, G.J.~Good and G.L.~Kane, {\it ibid.}~D {\bf 59}, 
   115004-1 (1999); S.~Pokorski, J.~Rosiek and C.A.~Savoy, hep--ph/9906206.
\bibitem{SYCHOI} S.Y.~Choi {\it et al.}, Eur.~Phys.~J.~C {\bf 7}, 
   123 (1999); G.~Moortgat--Pick and H.~Fraas, Phys.~Rev.~D {\bf 59}, 
   015016-1 (1998); hep--ph/9903220.
\bibitem{BK} M.~Brhlik and G.L.~Kane, Phys.~Lett.~B {\bf 437}, 331 (1998); 
   S.Y.~Choi, J.S.~Shim, H.S.~Song and W.Y.~Song, {\it ibid.}~B {\bf 449}, 
   207 (1999).
\bibitem{FO} T.~Falk and K.A.~Olive, hep-ph/9806236; T.~Falk, A.~Ferstl
   and K.A.~Olive, Phys. Rev. D {\bf 59}, 055009--1 (1999);
   hep-ph/9908311.
\bibitem{PW} A.~Pilaftsis and C.E.M.~Wagner, hep-ph/9902371;
   D.A.~Demir, hep-ph/9901389; J.F.~Gunion, B.~Grzadkowski, H.E.~Haber, 
   and J.~Kalinowski, Phys. Rev. Lett. {\bf 79}, 982 (1997); 
   B.~Grzadkowski, J.F.~Gunion and J.~Kalinowski, hep-ph/9902308;
   A.~M\'{en}dez and A.~Pomarol, Phys. Lett. {\bf B272}, 313 (1991); 
   A. Pilaftsis, hep-ph/9908373.
\bibitem{Ko} G.C.~ Branco, G.C.~Cho, Y.~Kizukuri and N.~Oshimo, 
   Phys.~Lett.~B {\bf 337}, 316 (1994); Nucl.~Phys.~{\bf B 449}, 
   483 (1995); D.A.~Demir, A. Masiero and O. Vives, 
   Phys.~Rev.~ Lett.~ {\bf 82}, 2447 (1999); 
   Y. G. Kim, P. Ko and J. S. Lee, Nucl. Phys. {\bf B 544}, 64 (1999)
   and references therein;
   S.W.~Baek and P.~Ko, Phys. Rev. Lett. {\bf 83}, 488 (1999);
   S.W.~Baek and P.~Ko, hep-ph/9904283.
\bibitem{CL} S.Y. Choi and J.S. Lee, hep-ph/9907496.
\bibitem{DEMIR} D.A. Demir, hep-ph/9905571. 
\bibitem{GDYN} see e.g. J.R. Primack, in {\it Dark Matter in Astro- and
   Particle Physics}, eds. H.V. Klapdor-Kleingrothaus and Y. Ramachers,
   (World Scientific, Singapore, 1997) p. 97; and astro-ph/9707285. 
\bibitem{BBN} see e.g. K.A. Olive and D.N. Schramm, in the Reviews of
   Particle Properties, Eur. Phys. J., {\bf C3} (1998) 1. 
\bibitem{JKG} G. Jungman, M. Kamionkowski, and K. Griest, Phys. Rep.
   {\bf 267} (1996) 195 and references therein.
\bibitem{FOS} T.~Falk and K.A.~Olive, Phys.~Lett.~B {\bf 375}, 196 (1996); 
   T.~Falk, K.A.~Olive and M.~Srednicki, {\it ibid.}~B {\bf 354}, 99 (1995).
\bibitem{Pilaftsis} D. Chang, W.-Y. Keung and A. Pilaftsis,
   Phys. Rev. Lett. {\bf 82}, 900 (1999).
\bibitem{HIGGS} see. e.g. L3 Collaboration, CERN Preprint CERN-EP98-72
   (1998).
\bibitem{G} K. Griest, Phys. Rev. {\bf D38} (1988) 2357.
\bibitem{DGH}M. Dugan, B. Grinstein and L. Hall, Nucl. Phys. {\bf B255}, 413
   (1985).
\bibitem{HH} H.E.~Haber and R.~Hempfling, Phys. Rev. D {\bf 48}, 4280 
   (1993); M.~Carena, J.R.~Espinoza, M.~Quir\'{o}s, and C.E.M.~Wagner,
   Phys. Lett. {\bf B355}, 249 (1995).
\bibitem{PILL}A. Pilaftsis,  Phys. Rev. {\bf D58} (1998) 096010, and 
    Phys. Lett. {\bf B435} (1998) 88.
\bibitem{DM} N.G.~Deshpande and E.~Ma, Phys. Rev. D {\bf 16}, 1583 (1977);
   Phys. Rev. D {\bf 18}, 2574 (1978).
\bibitem{CKCK} D.~Chang and W.-Y.~Keung and I.~Pillips, Phys. Rev. D 
   {\bf 48}, 3225 (1993); A.~Pilaftsis, Phys. Rev. Lett. {\bf 77}, 4996 
   (1996); S.Y.~Choi and M.~Drees, Phys. Rev. Lett. {\bf 81}, 5509 (1998);
   D.~Atwood and A.~Soni, Phys. Rev. D {\bf 52}, 6271 (1995).
\bibitem{EF} J. Ellis and R. Flores, Phys. Lett. {\bf B300}
   (1993) 175.
\bibitem{SMC} D. Adams et al. (The Spin Muon Collaboration), 
   Phys. Lett. {\bf B329} (1994) 399.
\bibitem{SVZ} M. A. Shifman, A. I. Vainshtein, and V. I. Zakharov, 
   Phys. Lett. {\bf B78} (1978) 443; A. I. Vainshtein,  V. I. Zakharov, 
   and M. A. Shifman Usp. Fiz. Nauk {\bf 130} (1980) 537. 
\bibitem{GLS}J. Gasser, H. Leutwyler, and M. E. Sainio, 
   Phys. Lett. {\bf B253} (1991) 252.
\bibitem{DN1}M. Drees and M. M. Nojiri, Phys.Rev. {\bf D47} (1993) 4226. 
\bibitem{DN2}M. Drees and M. M. Nojiri, Phys.Rev. {\bf D48} (1993) 3483. 
\bibitem{PDG} Particle Data Group, Euro. Phys. J. {\bf C3} (1998) 1.
\bibitem{GF} P.~Gondolo and K.~Freese, hep--ph/9908390.
\end{references}
\end{document}